\documentclass[11pt]{article}

\usepackage{geometry}
\usepackage{currfile}
\usepackage{textcomp}

\usepackage{xcolor,soul}
\sethlcolor{yellow}

\usepackage[pdftex]{graphicx}
\DeclareGraphicsExtensions{.pdf,.jpeg,.png}
\graphicspath{{./figures}}

\date{November 21, 2019}

\begin{document}

\title{\Large{Empowering Artists, Songwriters \& Musicians in a Data Cooperative \\
through Blockchains and Smart Contracts}\\
~~\\
\large{(Extended Abstract)}  \\
~~}
\author{
\large{Thomas Hardjono and Alex Pentland}\\
\large{~~}\\
\large{MIT Connection Science}\\
\large{Massachusetts Institute of Technology}\\
\large{Cambridge, MA, USA}\\
\large{~~}\\
\small{{\tt hardjono@mit.edu}~~ ~~{\tt pentland@mit.edu}}\\
\large{~~}\\
}

\maketitle

\begin{abstract}

Over the last decade there has been a continuing decline in social trust on the part of individuals with regards to the handling and fair use of personal data, digital assets and other related rights in general. At the same time, there has been a change in the employment patterns for many people through the emergence of the gig economy. These gig workers include artists, songwriters and musicians in the music industry. We discuss the notion of the data cooperative with fiduciary responsibilities to its members, which is similar in purpose to credit unions in the financial sector. A data cooperative for artists and musicians allows the community to share IT resources, such as data storage, analytics processing, blockchains and distributed ledgers. A cooperative can also employ smart contracts to remedy the various challenges currently faced by the music industry with regards to the license tracking management.
~~\\
~~\\
\end{abstract}

\newpage
\clearpage

\section{Introduction}

During the last decade, all segments of society have become increasingly 
alarmed by the amount of data, and resulting power, held by a small number of actors~\cite{PentlandHardjono2019a}.  
Data is, by some, famously called ``the new oil''~\cite{WEF2011} and comes 
from records of the behavior of citizens. 
Why then, is control of this powerful new  resource concentrated in so few hands?   
During the last 150 years, questions about concentration of power 
have emerged each time the economy has shifted to a new paradigm; 
industrial employment replacing agricultural employment, consumer banking replacing cash and barter, 
and now ultra-efficient digital businesses 
replacing traditional physical businesses and civic systems.   

As the economy was transformed by industrialization and then by consumer banking, 
powerful new players such as Standard Oil, J.P. Morgan, and a handful of others 
threatened the freedom of citizens. 
In order to provide a counterweight to these new powers, 
citizens joined together to form trade unions and 
cooperative banking institutions, which were federally chartered 
to represent their members' interests. 
These citizen organizations helped balance 
the economic and social power between 
large and small players and between employers and worker.

The same collective organization is required to move 
from an individualized asset-based understanding of data control 
to a collective system based on rights and accountability, 
with legal standards upheld by a new class of representatives 
who act as fiduciaries for their members.  
In the U.S. almost 100 million people are members of credit unions, 
not-for-profit institutions owned by their members, 
and already chartered to securely manage their members'
digital data and to represent them in a wide variety of financial transactions, 
including insurance, investments, and benefits. 
The question then is, could we apply the same push for citizen power 
to the area of data rights in the ever-growing digital economy?   

Indeed, with advanced computing technologies it is practically possible 
to automatically record and organize all the data that 
citizens knowingly or unknowingly give to companies and the government, 
and to store these data in credit union vaults.
In addition, almost all credit unions already manage their 
accounts through regional associations that use common software, 
so widespread deployment of data cooperative capabilities 
could become surprisingly quick and easy.

\section{Empowerment of Individuals: The Data Cooperative}
\label{sec:DataCoopIntro}

Over the last decade there has been a continuing decline in trust on the part of individuals
with regards to the handling and fair use of personal data~\cite{WEF2011,WEF2014}.
The public is increasingly aware of the power of big data
combined with advanced analytics and artificial intelligence.
They are also increasingly aware of the power wielded by social media platforms
in influencing their daily lives,
ranging from influencing the types of advertisements they see on websites and in the media,
to the types of good and services they purchase online.
Pew Research reported that 91 percent
of Americans agree or strongly agree that consumers 
have lost control over how personal data is collected and used, 
while 80 percent who use social networking sites are 
concerned about third parties accessing their shared data~\cite{Madden2014}.

The World Economic Forum (WEF) has already reported 
on the state of declining trust in its 2014 report on personal data~\cite{WEF2014},
and recommended ways to remedy the situation.
These recommendations include (i) increase in {\em transparency}
by providing individuals with insight and meaningful control,
(ii) improve {\em accountability} by orienting throughout the value
chain (front-end to back-end) with risks being equitably distributed,
and
(iii) {\em empowerment of individuals} by way giving them a say in how
data about them is used by organizations and 
by giving individuals the capacity
to use data for their own purposes.

At the same time as this decreasing trust with regards to the handling and fair use of personal data,
there has been a change in the employment patterns of many people through the
emergence of the {\em gig economy}
characterized by non-traditional, independent, short-term working relationships.
In some cases, new technological platforms
have paved the way and enabled the emergence of new kinds of 
gig employment previously unavailable (e.g. car ride-sharing services).
However,
the same technological advances that enabled the emergence of new forms of gig employment
may also hold individual participants as captive and dependent for their livelihoods
on these gig-enabling platforms.

\begin{figure}[!t]
\centering
\includegraphics[width=1.0\textwidth, trim={0.0cm 0.0cm 0.0cm 0.0cm}, clip]{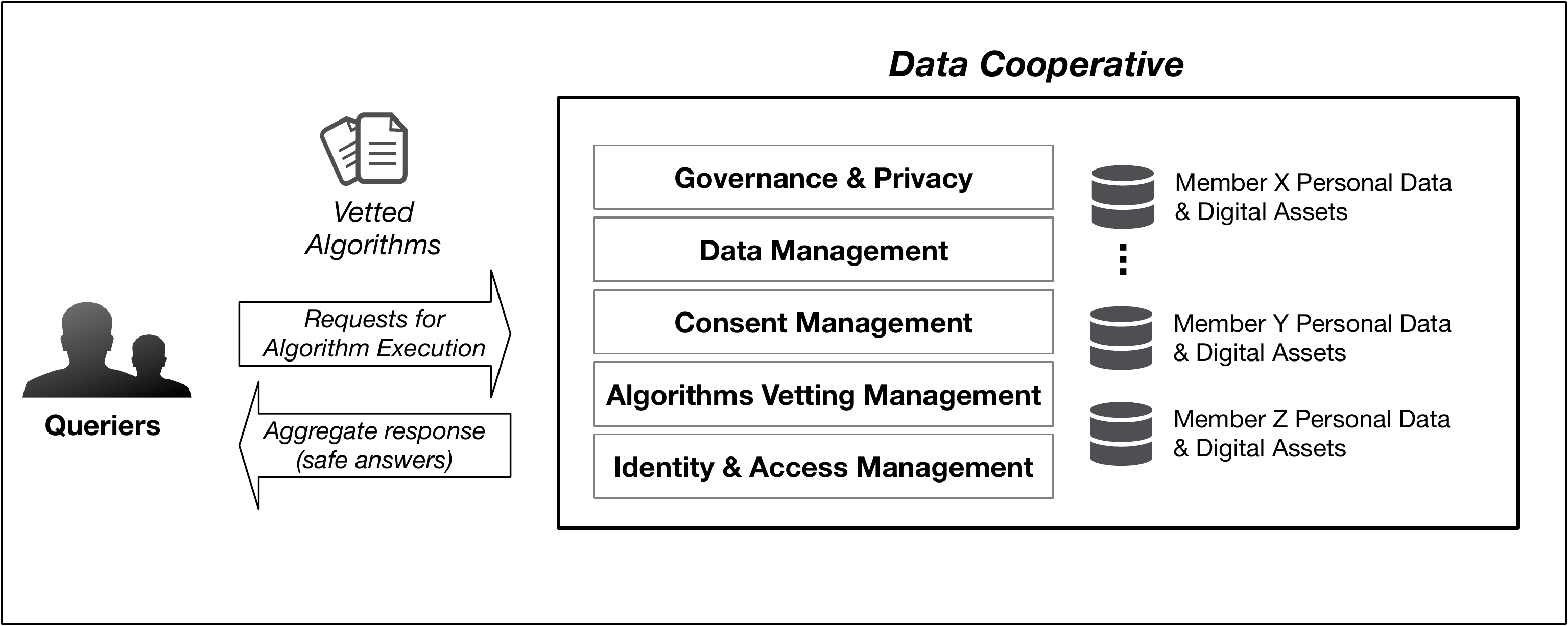}
	%
	%
\caption{Overview of the Data Cooperative based on the MIT Open Algorithms principles}
\label{fig:data-coop-ecosystem}
\end{figure}

Our notion of a digital {\em data cooperative} follows from the recommendation
of the 2014 WEF report.
We define the data cooperative as a {\em member-owned} organization that has a legal {\em fiduciary}
responsibility to its members in the access, management and use of the members' personal data
for their benefit.
In its simplest form the data cooperative could be one where
a group of individuals
with a common purpose voluntarily pool together 
access to their personal data, digital assets and other rights.

The overall goal of the member-owned data cooperative is
to empower the members as a collective by:
\begin{enumerate}

\item	{\em Shared insights through vetted algorithmic computations}:
By pooling together access to personal data and running algorithms
on the pooled data,
members can obtain insights about a particular aspect of the work or their lives.

Thus, for example, a cooperative for ride-sharing drivers 
may provide members with insights about their workday patterns for a given city or region.
A cooperative for artists and musicians
may allow the members to obtain aggregate insights related to incomes
in different continents of the world.

\item	{\em Shared computational IT platform}:
Having a common IT infrastructure for data management
and algorithmic computations,
allows the members obtain comparative insights 
that they could not otherwise obtain as a sole individual.

\item	{\em Providing collective leverage}:
By understanding aspects about their work or lives,
members become aware of the potential leverage they
have as a collective in dealing with changes and evolution
in the job market.
This may extend to leverage in obtaining group access and discounts 
to goods and services obtained through the cooperative.

\item	{\em Shared governance and privacy policies}:
A data cooperative is owned by its members,
and as such members have a say in the governance of the cooperative,
which includes actions permitted on the personal data of its members.
The governance rules must include policies
regarding the establishment of a set of approved algorithms
that can be executed on the data of the members.
The privacy of members must be foremost in the context the governance.
From a legal perspective,
the cooperative has a {\em fiduciary} obligation~\cite{Balkin2016} to its members,
akin to a credit union having a financial fiduciary relationship with its members.

\end{enumerate}

There is a range of things that members within a data cooperative can share with each other.
We refer to these broadly as ``personal data'',
which may may range from data generated by electronic devices (e.g. location data)
as a by-product of using the device,
data generated from using a 3rd party services (e.g. telephone call data records),
to biomedical data unique to a person (e.g. DNA sequence, etc.).
Data may also include those generated as a result of specific types of work
(e.g. hospital schedule sheet for a nurse) or 
produced by the work itself (e.g. compositions by a songwriter, number of views on social media, etc.).
The exact data being shared depends on the nature and purpose of the cooperative
and must be defined by the members of the cooperative.
The cooperative may also provide digital identity management
to its membership~\cite{HardjonoPentland2019c}.

Members of the cooperative may store their personal data and digital assets
at the data cooperative (e.g. in its cloud infrastructure).
Alternatively, they can store these elsewhere (e.g. in personal data stores)
and make a copy remotely accessible to the cooperative~\cite{HardjonoPentland2019d}.
Being a member-owned and voluntary organization,
an individual person is free at any time to leave the cooperative
and remove their personal data and other assets from the cooperative. 
A member retains legal ownership rights over their data and digital assets.

Our approach to data privacy -- referred to as the MIT Open Algorithms (OPAL)~\cite{HardjonoPentland2019d} --
consists of the development of several fundamental principles
that stem from the GDPR privacy requirements. 
The first principle is that data must never leave it repository.
Instead it is the algorithm that must be secure transmitted to the data repository and be executed there.
This means that data must never be exported from the IT infrastructure of the data cooperative.
Second,
only pre-approved algorithms that have been vetted to be fair and unbiased should be executed on the members data.
Third,
only coarse-granularity results (aggregate level) must be reported,
in order to prevent re-identification of individuals and thus protect their privacy.
Finally,
a robust consent management protocol must be used which permits the logging and auditing
of access to data and the executions of algorithms.

\section{Use Case: Empowering Artists \& Musicians}
\label{sec:ArtistsGigEconomy}

The music industry in the United States
today represents one of the most complex business ecosystems -- one that
is currently facing a number of challenges impacting
all the entities in the music supply chain~\cite{Fisher2004}.
These challenges are due to a number of factors,
including the emergence of new delivery mechanism (e.g. digital music streaming),
the change in listening habits of the younger generation of audiences,
the changing notions of music ownership by consumers,
the complex legal arrangements around music copyright,
and other factors.

Today
artists and musicians are part of the gig economy and function much as gig-workers.
They are finding it increasingly difficult to make a sustainable living
as artistic creators~\cite{Howard2017a,Howard2017}.
Many feel they lack visibility into state of license issuances of their works,
and therefore to the projected incomes.
For many songwriters and musicians,
royalty payments for licenses may take several months to arrive.
In many cases, the actual songwriters or artists for a given musical work
are considered to be ``un-attributeable'' -- which leads
to payments languishing
in escrow accounts around the world unclaimed by the songwriters to whom the payments are due~\cite{Messitte2015}.

The industry itself as a whole has not invested sufficiently in technological advances 
(e.g. digital delivery, cloud services, digital identity, etc),
and has in fact turned down various new opportunities related to digital music over the past two decades
(e.g. Naspter case in the late 1990s~\cite{Fisher2004}).
The net result is a music ecosystem today that still uses
outdated accounting setups (e.g. exchanging Excel spreadsheets),
leading to gross inefficiencies,
confusing royalty statements and delayed payments,
with music being tagged incorrectly leading to mistakes in attributions~\cite{Howard2017a}.

\begin{figure}[!t]
\centering
\includegraphics[width=1.0\textwidth, trim={0.0cm 0.0cm 0.0cm 0.0cm}, clip]{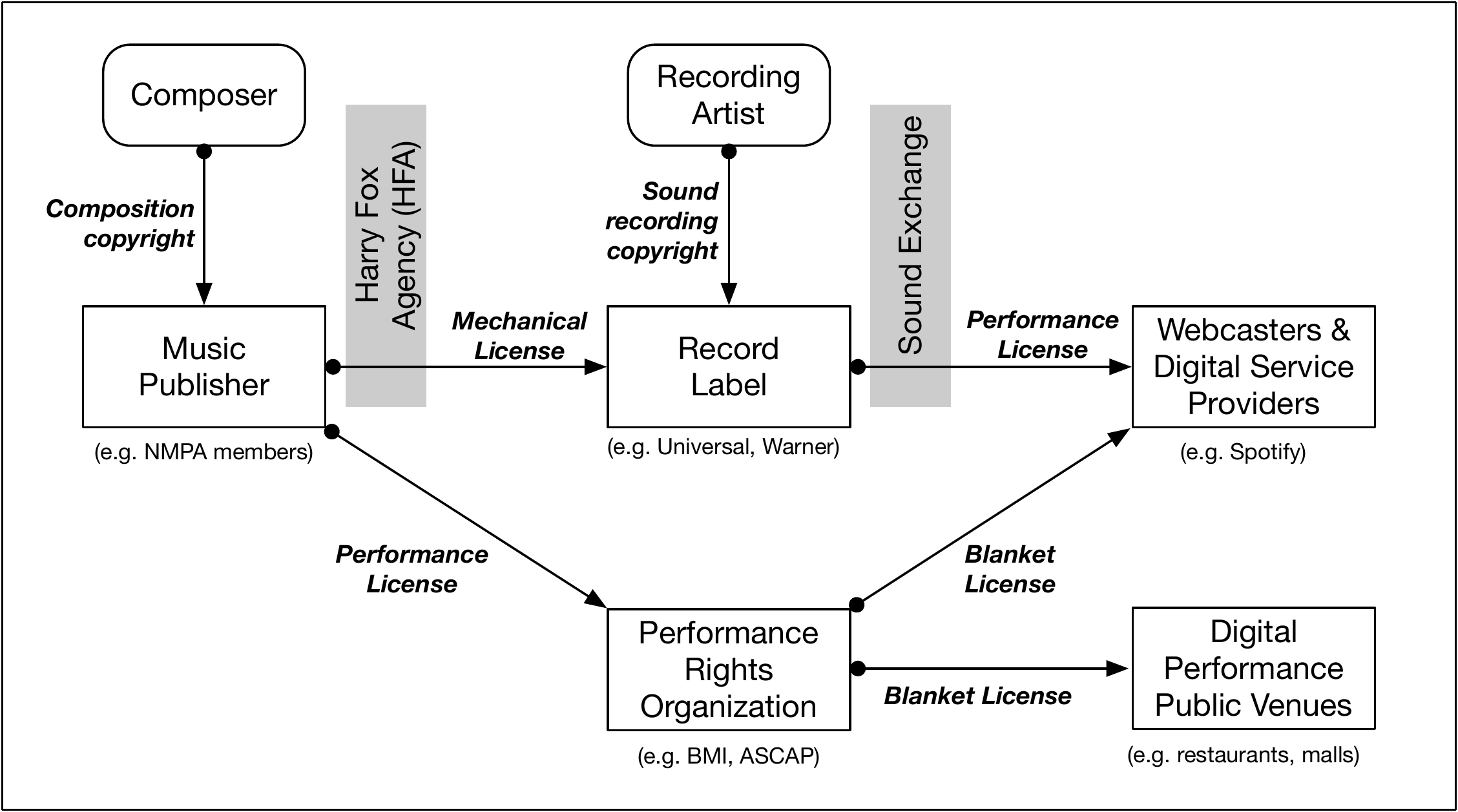}
	%
	%
\caption{Overview of the digital music licensing supply-chain}
\label{fig:supplychain}
\end{figure}

A comprehensive discussion of the music industry is beyond the scope of the current
work and has been treated elsewhere (e.g. see~\cite{Fisher2004,Passman2015}).
However, in order facilitate the discussion, 
we provide a high level summary of the roles and tasks
of the most common entities in the music license supply chain (see Figure~\ref{fig:supplychain}).
When a {\em composer} or songwriter creates a {\em musical work}, 
such as a music composition or score, they obtain copyright as soon as
the musical work is realized (e.g. transcribed on a piece of paper).
In order to facilitate the licensing of the composition,
the songwriter typically engages a {\em music publisher}
who may additionally manage the business relationship (e.g. manage contracts)
on behalf of the songwriter.
When a record label seeks to create a sound recording
of the composition performed by a recording artist,
the record label must obtain a {\em mechanical license} from the music publisher
(or directly from the songwriter).
The term ``mechanical'' derives from the days of the use of a physical devices (tape roll)
or physical mediums (e.g. LPs or compact discs) as the primary means of 
making the sound recordings available to consumers.

Similarly, when a music streaming provider (e.g. Spotify, Pandora) -- referred to more formally as 
{\em digital service providers} (DSP) -- seek to offers streaming services to consumers,
they must first obtain the appropriate {\em performance license} from the record labels 
who owns the legal rights to the sound recordings.
In order to collect and disburse royalties generated from digital performances (e.g. streaming)
the US Congress established a non-profit collective rights management organization
called {\em SoundExchange} (SE) in 2003.
One of the key roles of SoundExchange is to set the royalty rates for digital performances.
On the music publishing side,
an organization called the {\em Harry Fox Agency} (HFA) that was founded in 1927 has the task of managing,
collecting and disbursing the mechanical license royalty-fees on behalf of the music publishers
in the United States.

The Berklee College of Music and MIT have been
spearheading an effort called {\em Open Music} to explore new technological means
and new incentives mechanisms 
to enable a new open music ecosystem to evolve.
One outcome of this joint effort has been the formation of the 
Open Music Initiative (OMI)~\cite{PanayPentland2018} in 2017
as a forum for discussion regarding the various
aspects -- technological, business, employment models -- of the future music industry globally.

\section{A Data Cooperative for Artists \& Musicians}
\label{sec:RightsLedger}

We believe that artists and musicians as gig-workers
would do well in forming communities based on the notion of the data cooperative.
In the following we discuss some aspects of such a data cooperative.

\begin{figure}[!t]
\centering
\includegraphics[width=0.9\textwidth, trim={0.0cm 0.0cm 0.0cm 0.0cm}, clip]{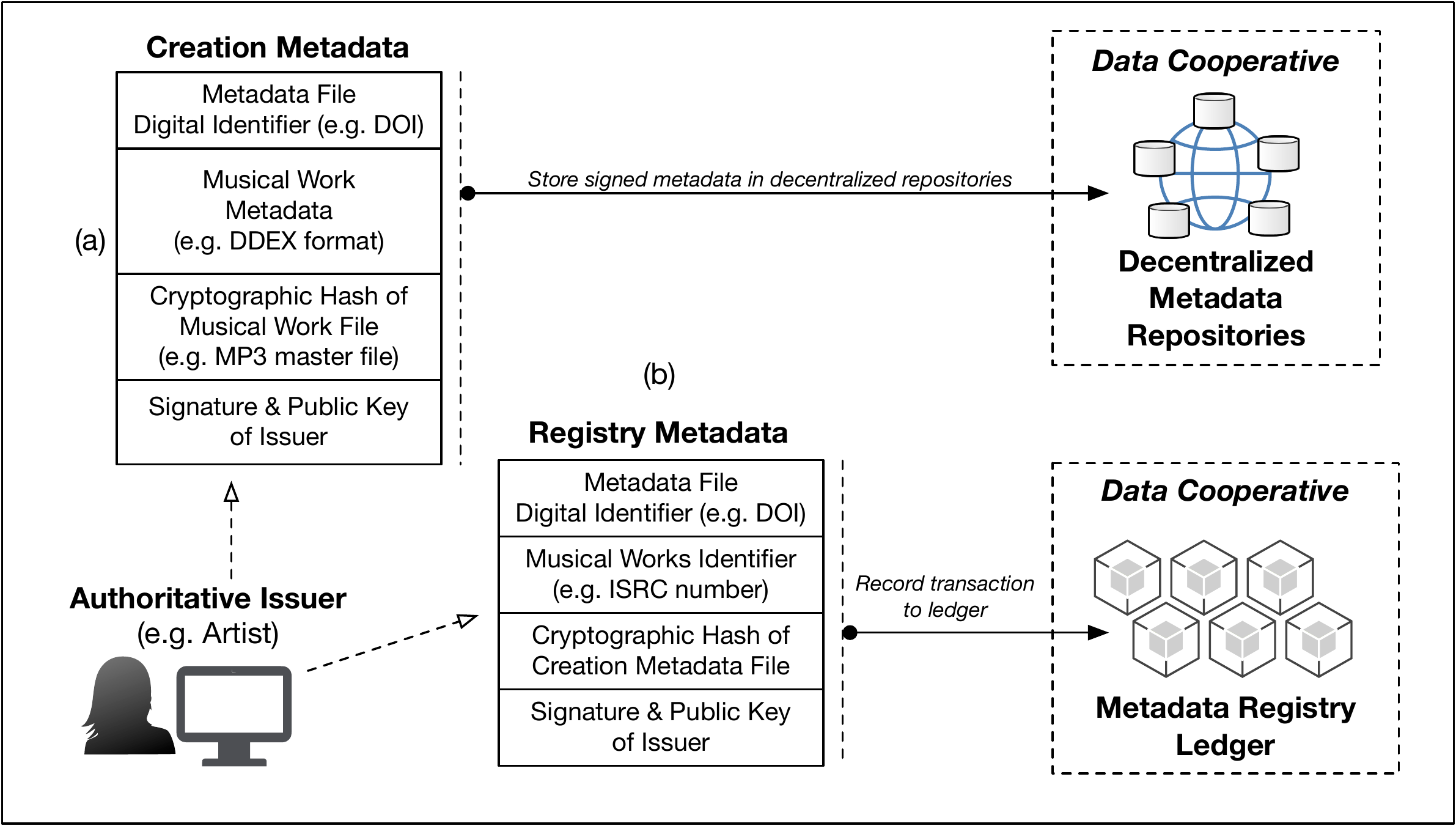}
	%
	%
\caption{The data cooperative distributed ledger for metadata management}
\label{fig:MetadataLayerComponents}
\end{figure}

\subsection{Shared Repositories for Musical Works and Artists Personal Data}

One of the significant issues in the music supply chain today 
is the lack of consistent, complete and authoritative 
information or metadata regarding the creation of 
a given {\em musical work} -- namely the individual composition or a sound recording track. 
In many cases multiple entities in the music supply chain have 
each created their own version of the metadata for a musical work, 
often by manually re-entering the same information or through 
scraping data from other sites~\cite{Deahl2019}. 
In such cases, the effort to synchronize or to correct 
the information becomes manually laborious and error-prone. 
Furthermore, confidential information regarding the legal ownership of 
the musical work is often commingled in the same metadata, 
making the entire database proprietary and thus closed. 
Currently the music industry have created standards for metadata file formats (e.g. DDEX based on XML),
but the industry as a whole does not as yet have widely adopted standards which define
the processes or workflows by which creation metadata is collected, displayed and validated.
This lack of standards for metadata workflows is only one of the many problems
plaguing the industry as a whole
(e.g. see~\cite{Howard2017a,Howard2017,Messitte2015}).

We believe the music industry needs to move to an alternative model 
for creation metadata following the open access paradigm found in other industries, 
such as in book publishing, library systems and in the automotive parts supply chain. 
Creation-metadata needs to be separated from rights-ownership metadata
as well as from licensing-metadata~\cite{HardjonoHoward2019}.
The creation metadata must not include the actual musical work itself (e.g. sound recording MP3 file)
and must not carry the legal ownership or copyright information
of the musical work.

The notion of a data cooperative is appealing here
as means to help communities of artists and musicians
in managing their musical works,
including the creation of authoritative metadata on a shared IT infrastructure.
This allows its members to manage their metadata files and 
musical works (e.g. MP3 master file; song composition files) at a lower cost 
while retaining control over these valuable assets.
This is illustrated in Figure~\ref{fig:MetadataLayerComponents}~(a).

\subsection{Shared Ledger for Music Metadata Registration}

Distributed ledger technology and blockchain systems~\cite{NIST-80202-2018} has captured
the attention of many artists and musicians in the past few years
as a potential new paradigm that may help them obtain a more accurate indicator regarding
the consumer adoption of their musical works
and provide them a more sustainable way of making a living
through a more direct transactions and payments cycle~\cite{Howard2017}.

A data cooperative for artists and musicians
can make available distributed ledgers to enhance and automate tasks or functions
related the music rights and licensing.
For example, the cooperative can establish
a {\em metadata registry ledger}
to which members can register their creation metadata.
The entries of the ledger 
includes a globally unique resolvable identifier that
allows anyone to follow the linked identifier to a copy of the complete
creation metadata somewhere on the Internet.
This is illustrated in Figure~\ref{fig:MetadataLayerComponents}~(b).

There are several possible benefits to such an approach.
The ``publishing'' of the signed registry-metadata file of a musical work
onto the registry ledger
provides legal support to the copyright claim on the part of the creator(s).
The distributed ledger as a whole acts as ``notarization'' service
where only the cooperative members have write-permission to add new entries to the ledger,
while anyone in the public can read the metadata and 
validate the digital signature via the ledger transaction entry.
This notarization using a distributed ledger
provides a relatively immutable and non-repudiable
timestamped public evidence of the existence of the musical work.

A second benefit of the metadata registry ledger
is that together with the metadata repository
it becomes the authoritative source of provenance information
regarding a given musical work.
By being the open-readable registry for musical works metadata,
the registry ledger effectively becomes the trusted source (or an ``oracle of truth'') for metadata
that can then be referenced (linked to) by other types of ledger-based transactions,
such as smart contracts that handle license issuance and rights-ownership exchanges.
Even existing systems (e.g. legacy databases)
can thus refer to the relevant entries (e.g. transaction-id) in the registry ledger.

\subsection{Smart Contracts for Music License Management}

Artists and musicians see
distributed ledgers and smart contracts as a promising avenue 
for a more direct transaction engagement model.
Smart contracts as stored procedures or functions (i.e. code)
available on the nodes of the P2P network
offers a number promising capabilities for increasing the efficiency
of business workflows.
In the context of the music contracts supply chain management,
there are several possible applications of smart contracts
implementing different business logic associated with different phases of the contracts supply chain.
Thus, for example, smart contracts could be used to implement the logic of the following tasks:
(i) the licensing of music copyrighted works
(e.g. performance license and mechanical license);
(ii) the tracking of payments for granted licenses;
(iii) the disbursement of royalty payments to the correct rights-holders;
and
(iv) the revocation of granted licenses or automatic expiration.

A data cooperative for artists and musicians could help their membership
generally with authoring smart contracts for specific ledger systems.
This allows the community to standardize on the legal terms of the license,
leaving the pricing decision to each artist/musician.
The sharing of common ``templates'' of smart contracts provides a way
for artists to save on legal costs.
A second possible role for a data cooperative is to operate
a distributed ledger for its membership, and possibly for other data cooperatives.
This approach allows different data cooperatives around the world
to share the costs of operating the shared ledger.

\begin{figure}[!t]
\centering
\includegraphics[width=1.0\textwidth, trim={0.0cm 0.0cm 0.0cm 0.0cm}, clip]{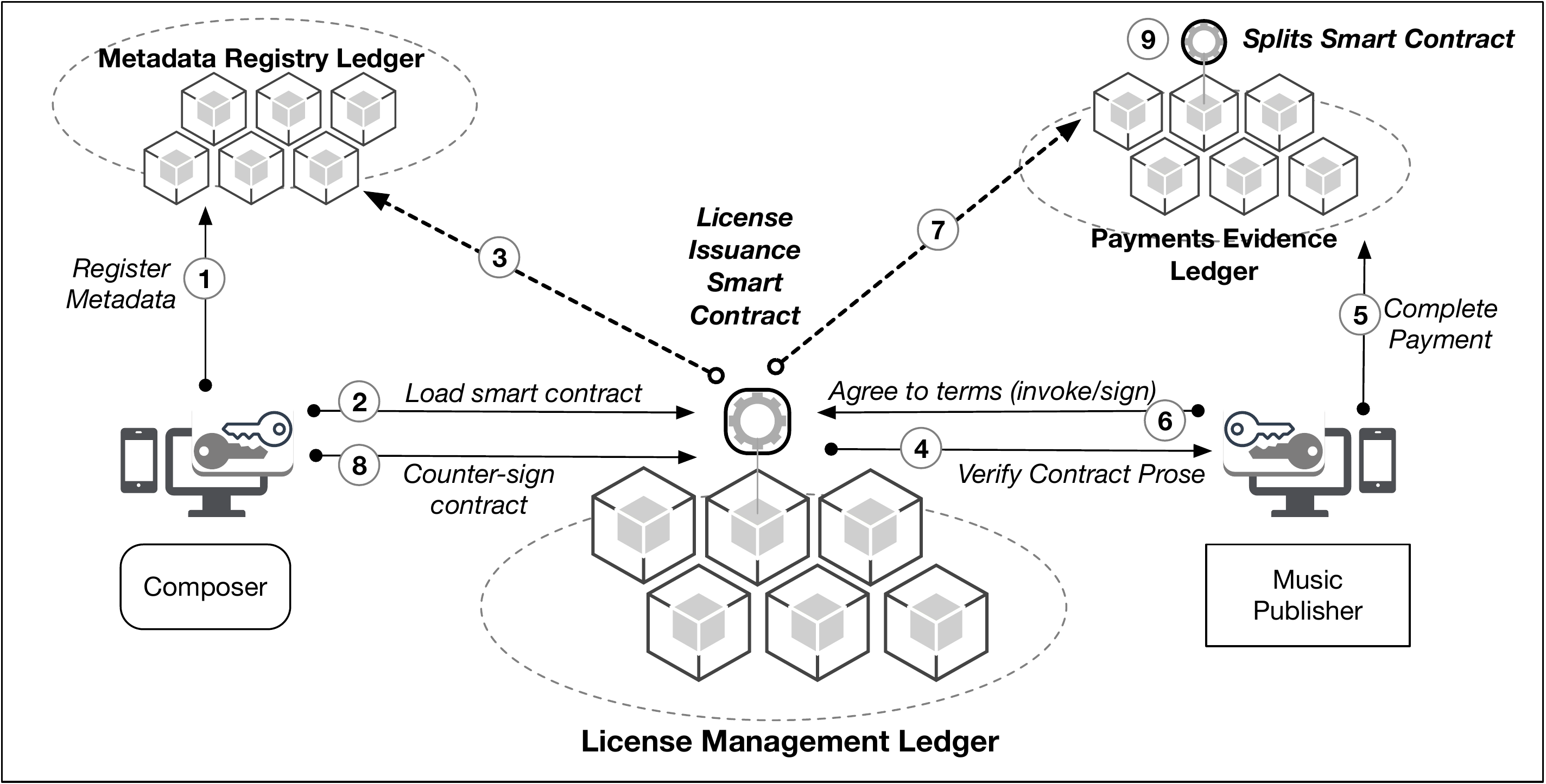}
	%
	%
\caption{Overview of a cooperative copyright license management ledger and smart contract}
\label{fig:blockchainlicensing}
\end{figure}

A sketch of this workflow 
and a copyright license management ledger
is shown in Figure~\ref{fig:blockchainlicensing}. 
Here, a composer or songwriter
employs a smart contract for the purpose of allowing
other entities (e.g. music publishers, other artists, etc.)
to obtain a license for the composer's musical work.
The composer must first record the metadata to the metadata registry ledger
in order to allow the smart contracts later to unambiguously refer to (point to) the musical work being licensed.
This provides precision of licensing in the case that several version of the
musical work exists (e.g. different length of sound recordings).
This is summarized in Steps~1 to 3 in Figure~\ref{fig:blockchainlicensing}.
In the simplest implementation,
the smart contract code could be one that incorporates the legal prose
of the license agreement (referred to as {\em Ricardian} smart contracts~\cite{Grigg2000} )
and where the code simply applies the digital signature of the licensee.

When a licensee (e.g. music publisher) seeks to obtain a copyright license
to a given composition,
the licensee needs to employ the correct smart contract on the
license management ledger.
Depending on the specific implementation of the smart contract,
the licensee may be required to make a payment in advance
(e.g. using separate payment mechanism)
and provide evidence to the smart contract regarding this payment.
This is summarized in Steps~4 to 7 in Figure~\ref{fig:blockchainlicensing}.
If a payments ledger is used,
then there is also the opportunity for a ``splits'' smart contract (Step~9)
to be used to automatically disburse the payment portions
to the correct rights holders in the case that
the copyright is jointly owned by multiple people
(e.g. composition created by multiple songwriters).

\section{Conclusions}
\label{sec:Conclusions}

Today we are in a situation where individual assets ...people's personal data... is being 
exploited without sufficient value being returned to the individual.  
This is analogous to the situation in the late 1800's and early 1900's 
that led to the creation of collective institutions 
such as credit unions and labor unions, 
and so the time seems ripe for the creation of 
collective institutions to represent the data rights of individuals. 

We have argued that data cooperatives with fiduciary obligations 
to members provide a promising direction for the empowerment 
of individuals through collective use of their own personal data. 
Not only can a data cooperative give the individual expert, 
community-based advice on how to manage, curate and protect 
access to their personal data, it can run internal analytics that 
benefit the collective membership. 
Such collective insights provide a powerful tool for negotiating 
better services and discounts for its members. 
Federally chartered credit unions are a useful model because they have already been legally 
empowered to act as cooperatives.
We we believe there are many other similar institutions that could 
also provide data cooperative services,
and we discuss the data cooperative model in the case
for empowering artists and musicians in the music industry.



\end{document}